%% file: arxiv.tex
\documentclass{article}
\usepackage{spconf,amsmath,amsfonts,graphicx}
\usepackage{bm}
\usepackage{booktabs}

\title{Blind and neural network-guided convolutional beamformer \\
for joint denoising, dereverberation, and source separation}
\name{Tomohiro Nakatani, Rintaro Ikeshita, Keisuke Kinoshita, Hiroshi Sawada, Shoko Araki}
\address{NTT Corporation, Japan\vspace{-2mm}}

\usepackage[absolute,showboxes]{textpos}
\TPMargin{5pt}
\newcommand{\copyrightstatement}{
\begin{textblock}{0.8}(0.1,0.01)
\noindent
\footnotesize
\copyright 2020 IEEE.  Personal use of this material is permitted.  Permission from IEEE must be obtained for all other uses, in any current or future media, including reprinting/republishing this material for advertising or promotional purposes, creating new collective works, for resale or redistribution to servers or lists, or reuse of any copyrighted component of this work in other works.
\end{textblock}
}
\begin{document}
\copyrightstatement
\input{macros.tex}

\ninept
\maketitle
\begin{abstract}
This paper proposes an approach for optimizing a Convolutional BeamFormer (CBF) that can jointly perform denoising (DN), dereverberation (DR), and source separation (SS).  First, we develop a blind CBF optimization algorithm that requires no prior information on the sources or the room acoustics, by extending a conventional joint DR and SS method. For making the optimization computationally tractable, we incorporate two techniques into the approach: the Source-Wise Factorization (SW-Fact) of a CBF and the Independent Vector Extraction (IVE). To further improve the performance, we develop a method that integrates a neural network (NN) based source power spectra estimation with CBF optimization by an inverse-Gamma prior. Experiments using noisy reverberant mixtures reveal that our proposed method with both blind and NN-guided scenarios greatly outperforms the conventional state-of-the-art NN-supported mask-based CBF in terms of the improvement in automatic speech recognition and signal distortion reduction performance.
\end{abstract}

\section{Introduction}
\label{sec:intro} 

When a speech signal is captured by distant microphones, e.g., in a conference room, it often contains reverberation, diffuse noise, and voices from extraneous speakers. They all reduce the intelligibility of the captured speech and often cause serious degradation in many speech applications, such as hands-free teleconferencing and Automatic Speech Recognition (ASR).

Recently, mask-based BeamFormers (BFs) \cite{souden10aslp,tranvu,BeamNET} have been actively studied to minimize the aforementioned detrimental effects in acquired signals. Masks indicate the time-frequency (TF) regions that are dominated by target speakers' voices and are used to estimate the acoustic transfer functions (ATFs) from the speakers to microphones. Many useful techniques have been proposed to estimate masks, e.g., by neural networks (NNs) \cite{BeamNET,hakan2015icassp} and clustering microphone array signals \cite{ito16eusipco,GSS}. The mask-based BF approach effectively optimizes BFs and Convolutional BFs (CBFs) that can jointly perform denoising (DN), dereverberation (DR), and source separation (SS) \cite{nakatani2020taslp,wangyou2020interspeech}. A drawback of this approach, however, is that ATFs and BFs are estimated based on different criteria, and thus the estimated ATFs are not guaranteed to be optimal for BF/CBF estimation.

Blind signal processing (BSP) \cite{common,ica,iva,hiroeiva,wpe,IP}, on the other hand, has been long studied for optimizing BFs/CBFs based only on observed signals with no prior information on source signals and room acoustics. Its performance can be further improved with power spectral estimates obtained using NNs \cite{Vincent2016dnnss,DNN-WPE,Kitamura2019TASLP,Kameoka2019NC}. A potential advantage of this approach over mask-based BFs/CBFs is that it can be optimized without relying on separate ATF estimation. However, this approach is limited because we have not yet developed effective and computationally efficient algorithms for jointly optimizing DN, DR, and SS (DN+DR+SS).

To overcome the above limitations, this paper first develops a new technique, called a blind CBF, that blindly estimates a CBF that can jointly perform DN+DR+SS in a computationally efficient way.  For this purpose, we extend a conventional CBF developed for blind DR+SS \cite{takuya2011taslp,Kagami2018icassp}. We empirically expect that this conventional CBF can also jointly perform DN when we use more microphones than target sources and separate stationary diffuse noise as additional sources. However, this extension greatly increases the computing cost as the number of microphones increases. To solve this problem, we incorporate two techniques that have recently been proposed for DR+SS and DN+SS: source-wise factorization (SW-Fact) of a CBF \cite{nakatani2020taslp,nakatani2020interspeech} and Independent Vector Extraction (IVE) \cite{koldovsky2019tsp,Robin2019waspaa,Ikeshita2020icassp}. 
Both respectively achieve computationally efficient optimization by factorizing a complicated multiple source DR step into simple single source DR steps based on weighted prediction error (WPE) \cite{wpe} and by omitting the separation of signals within a noise subspace. 
Although both techniques have been shown effective for their respective problems, their integration has not yet been investigated for DN+DR+SS.

This paper introduces two techniques that improve the estimation accuracy of the above extension. One is a coarse-fine source variance model for the blind CBF, which is shown to be indispensable by experiments for achieving high estimation accuracy.  Another is a NN-guided CBF, which incorporates source power spectra separately estimated by a NN into the blind CBF as an inverse-Gamma prior of the source variances.
Experiments using very challenging noisy reverberant speech mixtures show that our proposed blind CBF greatly outperforms the conventional state-of-the-art, a mask-based CBF \cite{nakatani2020taslp}, in terms of improvement in ASR performance and signal distortion reduction. With the NN-guided CBF, the performance can be further improved to a level that approaches one achieved by the conventional mask-based CBF with oracle mask information.

In the remainder of this paper, after a brief overview of related work in section 2, the problem formulation and proposed techniques are presented in sections 3 and 4. Experiments and concluding remarks are given in sections 5 and 6.

\section{Related work}
We also provide a comprehensive formulation of computationally efficient joint optimization for blind DN+DR+SS \cite{ikeshita2020ASJ,ikeshita2021icassp} by incorporating 
a CBF configuration \cite{ikeshita2019waspaa} into IVE \cite{Ikeshita2020icassp}. 
From that formulation, the algorithm proposed in this paper can be viewed as a variation using SW-Fact, and here we develop it by incorporating an IVE model into a CBF with SW-Fact \cite{nakatani2020interspeech}.  SW-Fact is a versatile technique with wide applications, including mask-based target speaker extraction \cite{nakatani2020taslp}, allowing us to use  computationally efficient optimization. In addition, this paper proposes techniques to improve the estimation accuracy of the proposed CBF (section \ref{sec:advance}).


Various techniques have integrated NNs and BSP \cite{Vincent2016dnnss,DNN-WPE,Kitamura2019TASLP,Kameoka2019NC}. Because these techniques directly use source power spectra estimated by NNs to update the coefficients of BFs/CBFs, NNs are required to present precise power spectral estimates. Although we could have chosen the same approach, our paper introduces looser integration of NNs and BSP using an inverse-Gamma prior, which makes overall optimization less sensitive to errors in the prior. Thus the framework can be easily extended to work with various prior information, such as speaker diarization in meetings \cite{GSS,nakatani17icassp}.

\section{Problem formulation}
Suppose that $J$ speech signals are captured by $M$ microphones with $M-J$ noise signals.\footnote{This assumption is introduced for algorithm derivation, and in practice the proposed method can perform DN even in diffuse noise environments, as shown by our experiments.}  This paper models the captured signals at each time $t$ and frequency $f$ in the short-time Fourier transformation domain as 
${\vect x}_{t,f}=\sum_{\tau=0}^{L_A-1}{\vect A}_{\tau,f}{\vect s}_{t-\tau,f}$,
where $\vect{x}_{t,f}=[x_{1,t,f},\ldots,x_{M,t,f}]^{\T}\in\mathbb{C}^{M\times 1}$ is a vector containing the captured signals, letting $(\cdot)^{\T}$ denote a non-conjugate transpose, $\vect{s}_{t,f}=[s_{t,f}^{(1)},\ldots,s_{t,f}^{(J)},s_{t,f}^{(J+1)},\ldots,s_{t,f}^{(M)}]^{\T}\in\mathbb{C}^{M\times 1}$ is a vector containing $J$ speech signals for $j\in[1,J]$ and $M-J$ noise signals for $j\in[J+1,M]$, and $\vect{A}_{\tau,f}\in\mathbb{C}^{M\times M}$ for $\tau=0,\ldots,L_A-1$ are convolutional transfer function matrices from the speech/noise sources to the microphones. 

For performing DN+DR+SS, we introduce a CBF:
\begin{equation}
\vect{y}_{t,f}=\vect{W}_{0,f}^{\HT}\vect{x}_{t,f}+\bar{\vect{W}}_{D,f}^{\HT}\bar{\vect{x}}_{t-D,f}.\label{eq:CBF}
\end{equation}
where $\vect{y}_{t,f}=[y_{t,f}^{(1)},\ldots,y_{t,f}^{(M)}]^{\top}$ is the CBF output, i.e., an estimate of $\vect{s}_{t,f}$, $\vect{W}_{0,f}\in\mathbb{C}^{M\times M}$ and $\bar{\vect{W}}_{D,f}\in\mathbb{C}^{M(L-D)\times M}$ are the coefficient matrices of the CBF, $\bar{\vect{x}}_{t-D,f}=[\vect{x}_{t-D,f}^\T,\ldots,\vect{x}_{t-L+1,f}^\T]^\T\in\mathbb{C}^{M(L-D)\times 1}$ is a vector containing a past captured signal sequence,
and $(\cdot)^{\HT}$ denotes a conjugate transpose. Prediction delay $D~(\ge 1)$ is introduced in Eq.~(\ref{eq:CBF}) to set the goal of dereverberation to reduce only the late reverberation and preserve the direct signal and early reflections \cite{wpe,kinoshita2009taslp}.

The optimization of the above CBF may be solved as a problem of blind DR+SS  \cite{takuya2011taslp}, where both the speech and noise sources are estimated as separate CBF outputs. However, the direct application of such an approach becomes computationally intractable as the number of microphones is increased. In addition, its effectiveness for DN has not been well investigated.

\begin{figure}
    \centering
    \includegraphics[width=0.801\columnwidth]{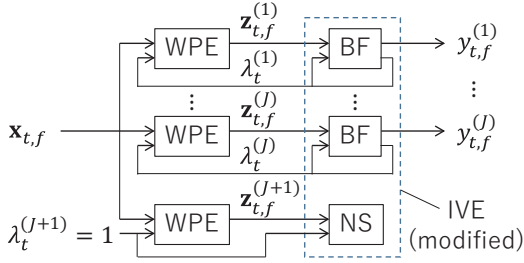}\vspace{-7mm}
    \caption{Processing flow of blind CBF with IVE source model: NS is a filter matrix that extracts mixture of noise (section~\ref{sec:IVE}).}\label{fig:flow}
    \label{fig:my_label}\vspace{-2mm}
\end{figure}

\section{Proposed method}
This section describes our proposed CBF that can achieve DN+DR+SS in a computationally efficient way. We first incorporate two techniques, the SW-Fact of a CBF used for DR+SS \cite{nakatani2020interspeech} and techniques used in IVE for DN+SS \cite{Robin2019waspaa,Ikeshita2020icassp}, into the proposed CBF in sections~\ref{sec:swfact}, \ref{sec:ivemodel}, and \ref{sec:optimization}. Then we introduce techniques that further improve CBF's estimation accuracy in section~\ref{sec:advance}.

\subsection{SW-Fact of a CBF}\label{sec:swfact}

As shown in a previous work \cite{nakatani2020taslp} and illustrated in Fig.~\ref{fig:flow}, a CBF at each $f$ in Eq.~(\ref{eq:CBF}), consisting of $\{\vect{W}_{0,f},\bar{\vect{W}}_{D,f}\}_\tau$, can be strictly factorized into a set of sub-filter pairs, each of which estimates source $y_{t,f}^{(j)}$ and consists of a WPE filter with prediction matrix $\vect{G}_{f}^{(j)}\in\mathbb{C}^{M(L-D)\times M}$ in Eq.~(\ref{eq:swwpe}) and a BF $\mathbf{q}_{f}^{(j)}\in\mathbb{C}^{M\times 1}$ in Eq.~(\ref{eq:swsep}):
%
\begin{align}
\vect{z}_{t,f}^{(j)}&=\vect{x}_{t,f}-\left(\vect{G}_f^{(j)}\right)^{\HT}\OL{\vect{x}}_{t-D,f},\label{eq:swwpe}\\
y_{t,f}^{(j)}&=\left(\vect{q}_{f}^{(j)}\right)^{\HT}\vect{z}_{t,f}^{(j)}.\label{eq:swsep}
\end{align}
Here, Eq~(\ref{eq:swwpe}) dereverberates the $j$th source signal in $\vect{x}_{t,f}$ using $\vect{G}_{f}^{(j)}$ and 
$\bar{\vect{x}}_{t-D,f}$, yielding dereverberated signal $\vect{z}_{t,f}^{(j)}\in\mathbb{C}^{M\times1}$. Equation~(\ref{eq:swsep}) extracts the $j$th source signal from $\vect{z}_{t,f}^{(j)}$.  
Because each sub-filter pair in the factorized form separately estimates each source, this is called SW-Fact.

As shown in section~\ref{sec:wpe}, using SW-Fact is advantageous for reducing the computational cost of the optimization. 

\subsection{Probabilistic formulation with an IVE source model}\label{sec:ivemodel}
Next, following the ideas of IVE \cite{Robin2019waspaa,Ikeshita2020icassp}, we define a source model, hereafter referred to as an IVE source model, as
\begin{align}
    p(\{y_{t,f}^{(j)}\}_{t,f,j})&=\prod_{t,f,j} p(y_{t,f}^{(j)}),\label{eq:ica}\\
    p(y_{t,f}^{(j)})&=\left\{\begin{array}{l}
    {\cal N}(0,\lambda_{t}^{(j)})~~\mbox{for}~~j\in[1,J],\\
    {\cal N}(0,1),~~\mbox{for}~~j\in[J+1,M],\end{array}\right.
    \label{eq:smodel}
\end{align}
where ${\cal N}(0,\lambda)$ denotes a complex Gaussian distribution with a mean zero and variance $\lambda$. Equation~($\ref{eq:ica}$) specifies mutual independence between the sources. With this model, although speech signals are modeled as time-varying Gaussians with time-varying and frequency-independent variances $\lambda_{t}^{(j)}$, noise signals are assumed to be stationary Gaussians. Due to this model, we can omit separation between noise sources as will be shown in section~\ref{sec:IVE}. 

Then, similar to the conventional blind CBF for DR+SS \cite{takuya2011taslp,nakatani2020interspeech}, the likelihood function can be derived, based on Eqs.~(\ref{eq:CBF})-(\ref{eq:smodel}) as
\begin{align}
{\cal L}(\theta)=&
-\sum_{t,f}\left[\sum_{j\in[1,J]}\left(\log\lambda_{t}^{(j)}+\frac{|{y}_{t,f}^{(j)}|^2}{\lambda_{t}^{(j)}}\right)+\sum_{j\in[J+1,M]}|{y}_{t,f}^{(j)}|^2\right]
\nonumber\\
&+2T\sum_f\log|\det {\vect{Q}}_{f}|,\label{eq:likelihood}
\end{align}
where $\vect{Q}_f=[\vect{q}_f^{(1)},\ldots,\vect{q}_f^{(M)}]\in\mathbb{C}^{M\times M}$ is a separation matrix, $\theta=\{\Theta_{\vect{G}},\Theta_{\vect{Q}},\Theta_{\lambda}\}$ is a set of parameters to be estimated, letting ${\Theta_\vect{G}}=\{\{\vect{G}_{f}^{(j)}\}_f\}_j$, ${\Theta_\vect{Q}}=\{\vect{Q}_{f}\}_f$, and $\Theta_{\lambda}=\{\{\lambda_{t}^{(j)}\}_t\}_j$, and $T$ is the total number of time frames. 

\subsection{Optimization algorithm}\label{sec:optimization}
We optimize a CBF by finding $\theta$ that maximizes the above likelihood function. Since no closed form solutions are known for it, we use iterative estimation based on a coordinate ascent method \cite{takuya2011taslp}, where $\Theta_{\vect{G}}$, $\Theta_{\lambda}$, and $\Theta_{\vect{Q}}$ are alternately updated while fixing the others, and the updates are iterated until a convergence is obtained. In the following, we derive equations for each update, and show how the above models can contribute to the computational efficiency for performing DN+DR+SS. 

\subsubsection{Update of $\Theta_{\vect G}$}\label{sec:wpe}
By fixing $\Theta_{\lambda}$ and $\Theta_{\vect Q}$ at their previously updated values and ignoring the constant terms, the likelihood function for $\Theta_{\vect G}$ becomes
\begin{align}
{\cal L}(\Theta_{\vect G})
&=-\sum_{f,j\in[1,M]}\left\|\left({\vect{G}}_f^{(j)}-\left({\vect{R}}_{f}^{(j)}\right)^{-1}\vect{P}_{f}^{(j)}\right){\vect{q}}_f^{(j)}\right\|_{{\vect{R}}_{f}^{(j)}}^2,\label{eq:likelihoodG}
\end{align}
where $\left\|\vect{x}\right\|_{\vect R}^2=\vect{x}^{\HT}\vect{R}\vect{x}$, $\vect{R}_{f}^{(j)}=\sum_t{\bar{\vect{x}}_{t-D}\bar{\vect{x}}_{t-D}^{\HT}}/\lambda_{t}^{(j)}$, $\vect{P}_{f}^{(j)}=\sum_t{\bar{\vect{x}}_{t-D}\vect{x}_{t}^{\HT}}/\lambda_{t}^{(j)}$, and we set $\lambda_{t}^{(j)}=1$ for $j\in[J+1,M]$ (i.e., noise signals) for notation simplicity.
Then, Eq.~(\ref{eq:likelihoodG}) can be maximized, not depending on $\Theta_{\vect{Q}}$, by updating ${\vect{G}}_f^{(j)}$ for each $j$ as
\begin{align}
{\vect{G}}_f^{(j)}\leftarrow\left({\vect{R}}_{f}^{(j)}\right)^{-1}\vect{P}_{f}^{(j)}~~\mbox{for}~~j\in[1,J+1].\label{eq:gupdate}
\end{align}
Here a clear advantage of using the IVE model is that we can skip calculating $\vect{G}_{f}^{(j)}$ for $j\in[J+2,M]$; they take the same value as $\vect{G}_{f}^{(J+1)}$ because all the noise signals share identical variance $\lambda_{t}^{(j)}=1$. As an advantage of using SW-Fact, we can separately optimize $\vect{G}_{f}^{(j)}$ for each source \cite{nakatani2020interspeech}, which makes the size of the covariance matrix ${\vect{R}}_{f}^{(j)}$ (required for the update) $M$ times smaller than the conventional CBF for DR+SS \cite{takuya2011taslp,Kagami2018icassp}. This greatly reduces the computational cost.

\subsubsection{Update of $\Theta_{\vect Q}$}\label{sec:IVE}

By fixing $\Theta_{\lambda}$ and $\Theta_{\vect G}$ and ignoring the constant terms, the likelihood function for updating $\Theta_{\vect Q}$ becomes
\begin{align}
{\cal L}(\Theta_{\vect Q})=&-\sum_{f,j\in[1,M]}\left\|\vect{q}_f^{(j)}\right\|_{\Sigma_{f}^{(j)}}^2+2T\sum_f\log|\det\vect{Q}_f|,
\end{align}
where
$\Sigma_{f}^{(j)}=\sum_t{\vect{z}_{t,f}^{(j)}\left(\vect{z}_{t,f}^{(j)}\right)^{\HT}}/{{\lambda}_{t}^{(j)}}$ and $\vect{z}_{t,f}^{(j)}$ is a dereverberated signal obtained by Eq.~(\ref{eq:swwpe}) using ${\vect{G}}_f^{(j)}$.
Because the above likelihood function is equivalent to that of the conventional IVE \cite{Ikeshita2020icassp} except that covariance matrices $\Sigma_{f}^{(j)}$ are obtained by source dependent dereverberated signals $\vect{z}_{t,f}^{(j)}$, we can use the same algorithm proposed for the conventional approach with minor modifications. 

With IVE, the filters for separating speech signals, $\vect{q}_f^{(j)}$ for $j\in[1,J]$, are updated based on Iterative Projection (IP) \cite{IP} as
\begin{align}
{\vect{q}}_f^{(j)}&\leftarrow\left({\vect{Q}}_f^{\HT}\Sigma_{f}^{(j)}\right)^{-1}\vect{e}_j,\\
{\vect{q}}_f^{(j)}&\leftarrow{\vect{q}}_f^{(j)}/\left\|\vect{q}_f^{(j)}\right\|_{\Sigma_{f}^{(j)}},
\end{align}
where $\vect{e}_j$ is the $j$th column of identity matrix $\vect{I}_{M}\in\mathbb{R}^{M\times M}$. In contrast, because our interest is not noise separation, IVE \cite{Robin2019waspaa,Ikeshita2020icassp} only updates filter matrix  $\vect{Q}_{N,f}=[\vect{q}_f^{(J+1)},\ldots,\vect{q}_f^{(M)}]$ that extracts a mixture of noise without distinguishing noise sources by
\begin{align}
    {\vect{Q}}_{N,f} \leftarrow\left(\begin{array}{c}
    -({\vect{Q}}_{S,f}^{\HT}\Sigma_f^{(J+1)}\vect{E}_S)^{-1}({\vect{Q}}_{S,f}^{\HT}\Sigma_f^{(J+1)}\vect{E}_N)\\
    \vect{I}_{M-J}
    \end{array}\right),
\end{align}
where ${\vect{Q}}_{S,f}=[{\vect{q}}_f^{(1)},\ldots,{\vect{q}}_f^{(J)}]$, and $\vect{E}_S\in\mathbb{R}^{M\times J}$ and $\vect{E}_N\in\mathbb{R}^{M\times(M-J)}$ are the first $J$ and the remaining $M-J$ columns of $\vect{I}_M$. Since ${\vect{Q}}_{N,f}$ can be obtained by a single step update that does not depend on the number of microphones, we can largely reduce the computational cost especially when we have many microphones. 

\subsubsection{Update of $\Theta_{\vect \lambda}$}\label{sec:lambda}

For the update of $\Theta_{\lambda}$, the likelihood function becomes
${\cal L}(\Theta_{\lambda})=-\sum_{t,f,j\in[1,J]}\left(\log\lambda_{t}^{(j)}+{|y_{t,f}^{(j)}|^2}/{\lambda_{t}^{(j)}}\right)$, where $y_{t,f}^{(n)}$ is obtained by Eq.~(\ref{eq:swsep}). 
Letting $F$ be the number of frequency bins, we obtain
\begin{align}
{\lambda}_{t}^{(j)}&\leftarrow \frac{1}{F}\sum_{f=0}^{F-1}|y_{t,f}^{(j)}|^2~~\mbox{for}~~j\in[1,J].
\label{eq:coarselambda}
\end{align}
Because the IVE model uses frequency independent source variances $\lambda_t^{(j)}$, we can separate $s_{t,f}^{(j)}$ over all $f$ as independent vector \cite{iva} and solve the permutation re-alignment to some extent even without any post-processing \cite{sawada11aslp}.

\subsection{Advanced source models}\label{sec:advance}

As will be shown in our experiments, however, the IVE source model does not necessarily provide the best performance of this approach. In the following, we present two variations of source models that can improve the performance.

\subsubsection{Coarse-fine source variance model}
Although the IVE model is effective for SS to extract targets, it also degrades the frequency resolution of the source model. Our preliminary experiments revealed that it severely limits  DR's estimation accuracy. To overcome that limitation, we propose a hybrid approach, called a coarse-fine source variance model. 
While we adopt Eq.~(\ref{eq:coarselambda}) for updating the variances of SS, we use a frequency dependent source model for DR. With this model, source variances $\lambda_{t,f}^{(j)}$ take different values at different frequencies for DR, updated by
\begin{align}
{\lambda}_{t,f}^{(j)}&\leftarrow|y_{t,f}^{(j)}|^2~~\mbox{for}~~j\in[1,J],\label{eq:finelambda}
\end{align}
and used for the calculation of Eq.~(\ref{eq:gupdate}).

\subsubsection{NN-guided inverse-Gamma prior}
A powerful alternative is to incorporate a NN into a source model. We use the power spectra of each source, $\gamma_{t,f}^{(j)}$, estimated by a NN, to define a prior distribution of $\lambda_{t,f}^{(j)}$. We adopt inverse-Gamma distribution $IG(\lambda;\alpha,\beta)=\beta^{\alpha}\Gamma(\alpha)^{-1}\lambda^{-(\alpha+1)}\exp(-\beta/\lambda)$ as the conjugate prior of the Gaussian source model, and set $\beta=\gamma_{t,f}^{(j)}$. Then, $\lambda_{t,f}^{(j)}$ is updated based on maximum a posteriori estimation by
\begin{align}
    \lambda_{t,f}^{(j)}\leftarrow\frac{|y_{t,f}^{(j)}|^2+\gamma_{t,f}^{(j)}}{\alpha+2}~~\mbox{for}~~j\in[1,J],
\end{align}
and used to update both DR and SS. (Note that projection back \cite{projback} is applied before this update to avoid scale ambiguity.)

In experiments, we set $\alpha=1$, and adopted a convolutional NN (CNN) \cite{CNN-uPIT} with a large receptive field that resembles the one used by a fully-Convolutional Time-domain Audio Separation Network (Conv-TasNet) \cite{ConvTasnet2019TASLP}. We trained the CNN to estimate masks based on utterance-level permutation invariant training criterion, and obtained $\gamma_{t,f}^{(j)}$ by applying the estimated masks to the captured signals. 

\begin{table*}[!t]
\renewcommand{\arraystretch}{1.1}
\caption{FWSSNRs (dB), CDs (dB), and STOIs for REVERB-2MIX SimData, and WERs (\%) for RealData, achieved by different BFs and CBFs after 10/100 iterations of WPE/IVE optimization. Results obtained with fewer iterations are also shown for WERs.}\label{tbl:exp}\smallskip\small
\label{table_example}
\centering
\begin{tabular}{cccccccccccc}\toprule
BF/CBF & Estimation & Source& FWSSNR & CD & STOI && \multicolumn{5}{c}{WER}\\
\cline{4-6}\cline{8-12}
type & type & model & 10/100 & 10/100 & 10/100 && 2/20 & 4/40 & 6/60 & 8/80 & 10/100\\
\midrule
Obs & n/a & n/a & 1.12 & 5.44 & 0.55 && \multicolumn{5}{c}{62.49~~~ (\#Iterations: n/a)}
\\ \hline
IVE & Blind & IVE & 4.78 & 4.02 & 0.76 && 49.65 & 39.55 & 36.63 & 34.94 & 34.36 \\
WPE+IVE & Blind & Coarse-fine & 5.86 & 3.67 & 0.83 && 30.48 & 22.67 & 20.40 & 19.81 & 19.54\\
CBF (proposed) & Blind & IVE & 4.29  & 3.60 & 0.78 && 38.67 & 28.80 & 26.55 & 26.04 & 25.65\\
CBF (proposed) & Blind & Coarse-fine & {\bf 6.16}  & {\bf 3.48} & {\bf 0.84} && 
{\bf 29.42} & {\bf 18.34} & {\bf 16.55} & {\bf 16.40} & {\bf 16.31} \\ \hline
IVE & NN-guided & NN prior & 5.55 & 3.71 & 0.81 && 27.26  & 28.30 & 28.01 & 28.03 & 28.38\\
WPE+IVE  & NN-guided & NN prior & 6.61 & 3.36 & 0.88 && 15.67 & 15.88 & 16.60 & 16.02 & 16.69 \\
CBF (proposed) & NN-guided & NN prior & {\bf 7.37} & {\bf 3.17} & {\bf 0.92} && 
{\bf 14.33} & {\bf 14.33} & {\bf 14.03} & {\bf 13.71} & {\bf 13.24} \\
\hline
CBF \cite{nakatani2020taslp} & Mask-based 
& NN mask & 5.79 & 3.62 & 0.82 && 21.08 & 20.09 & 20.13 & 19.78 & 19.59\\
CBF \cite{nakatani2020taslp} & Mask-based 
& Oracle mask & 7.42 & 2.95 & 0.91 && 
12.43 & 12.30 & 12.40 & 12.36 & 12.53\\
CBF \cite{nakatani2020interspeech} & Blind & IVA & 4.85 & 3.27 & 0.81 && 30.70 & 29.28 & 28.93 & 28.97	& 28.30\\
\bottomrule
\end{tabular}
\end{table*}

\section{EXPERIMENTS}
\label{sec:experiments}

This section experimentally evaluates the performance of our proposed techniques in terms of signal distortion reduction and ASR performance improvement. Due to space limitation, we skip the evaluation of the computational efficiency of SW-Fact and IVE, which was firmly confirmed in previous papers \cite{nakatani2020interspeech,Ikeshita2020icassp}.

\subsection{Dataset, methods compared, and evaluation metrics}

For the evaluation, we used the REVERB-2MIX dataset \cite{reverb2mix}, which is composed of noisy reverberant speech mixtures. Each mixture was created by mixing two utterances (i.e., $J=2$) extracted from the REVERB Challenge dataset (REVERB) \cite{REVERB}. Following the REVERB-2MIX guideline, evaluation was performed using separated signals that correspond to the evaluation set in REVERB.

With the blind estimation scenario, we compared four different methods: IVE \cite{Ikeshita2020icassp}, a cascade configuration of WPE \cite{wpe} followed by IVE (WPE+IVE), and two variations of CBFs (proposed) with the IVE source model (IVE) and with the coarse-fine source variance model. With the NN-guided scenario, IVE, WPE+IVE, and CBF (proposed) were compared. 
The number of iterations for optimization was set at 10 for WPE and 100 for IVE in all the experiments. For example, we updated WPE once every 10 IVE updates in the optimization of the CBFs. We choose this scheme because the convergence of WPE is generally much faster than that of IVE and the computational cost of WPE per iteration is much larger than that of IVE.
For all the methods, we applied projection back \cite{projback} and permutation re-alignment \cite{sawada11aslp} post-processings because they always improved the performance.
We set the frame length and the shift to 32 and 8 ms. A Hann window was used for the short-time analysis. The sampling frequency was 16 kHz and $M = 8$ microphones were used. 
For WPE and CBF, the prediction delay was set to $D = 2$ and the prediction filter lengths were respectively set at $L = 20, 16$, and $8$ for frequency ranges of 0 to 0.8, 0.8 to 1.5, and 1.5 to 8 kHz. 

We evaluated the speech enhancement performance using SimData in the REVERB-2MIX based on objective measures \cite{metrics}, including the Frequency-Weighted Segmental SNR (FWSSNR), the Cepstrum Distance (CD), and {the Short-Time Objective Intelligibility measure (STOI) \cite{stoi}}. 
In addition, we evaluated the ASR scores of the separated utterances using RealData in REVERB-2MIX. For the evaluation, we used a baseline ASR system developed for REVERB with Kaldi \cite{kaldi} that was composed of a trigram language model, and a TDNN acoustic model trained using a lattice-free MMI and online i-vector extraction. They were trained on the REVERB training set. 

\subsection{Evaluation results}
\label{results}
Table \ref{tbl:exp} shows the evaluation results of all the methods as well as the values calculated on the observed signals (Obs).
Denoted by CBF \cite{nakatani2020taslp}, the table also shows previous results \cite{nakatani2020taslp} obtained using mask-based CBFs based on masks estimated by a NN and based on oracle masks. 
Denoted by CBF \cite{nakatani2020interspeech}, the table also shows the results obtained using the previously proposed CBF \cite{nakatani2020interspeech}. With CBF \cite{nakatani2020interspeech}, the number of sources to be separated is set equal to the number of microphones, i.e., $I=M$, and the CBF was optimized using an Independent Vector Analysis (IVA) source model, where all the sources are modeled by time-varying Gaussians.

As shown in the table, CBF (proposed) with the coarse-fine model and CBF (proposed) with the NN-prior respectively achieved the best scores on all the metrics in the blind and NN-guided scenarios.
On the other hand, CBF (proposed) with the IVE model underperformed WPE+IVE. That is, joint optimization using the IVE model degraded the performance from cascade optimization. This shows that the coarse-fine model was indispensable for making the optimization of a blind CBF useful in this experiment.

When we compared the proposed CBFs with the conventional mask-based CBFs \cite{nakatani2020taslp}, 
the blind CBF with the coarse-fine model largely outperformed the mask-based CBF with masks estimated by a NN. In addition, CBF with the NN-guided model achieved scores close to those obtained by the mask-based CBF with oracle masks. These results clearly show the superior effectiveness of the proposed CBF with both blind and NN-guided scenarios. 



Finally, when we compared the CBF (proposed) using the IVE source model and CBF \cite{nakatani2020interspeech} using the IVA source model, the former outperformed the latter for improving WERs and was comparable to the latter for improving signal distortion scores. This means that using the IVE source model was effective to improve the computational efficiency without degrading the estimation accuracy from the IVA source model.

\section{Concluding remarks}
\label{sec:ref}

This paper proposed a method for optimizing a CBF that can jointly perform DN+DR+SS.  First, we developed a computationally efficient blind CBF algorithm by incorporating two techniques into a conventional CBF algorithm for joint DR+SS: SW-Fact and IVE. Then, we further improved the CBF performance by presenting two advanced source models: a coarse-fine source variance model for blind estimation and an inverse-Gamma prior model for NN-guided estimation. 
Experiments using noisy reverberant mixtures showed that the proposed blind and NN-guided CBFs greatly outperformed the conventional state-of-the-art mask-based CBF. In particular, the NN-guided CBF achieved a WER (13.24 \%) that is close to the WER (12.53 \%) that can be obtained by the conventional mask-based CBF only with oracle mask information.

\footnotesize
\bibliographystyle{IEEEtran}
\bibliography{mybib2}

\end{document}

%% file: macros.tex
\newcommand{\vect}{\mathbf}
\newcommand{\HT}{\mathsf{H}}
\newcommand{\T}{\top}
\newcommand{\WPD}{ML-ConvBF}
\newcommand{\OL}{\overline}
\newcommand{\LOL}[1]{\underline{{#1}}}
\newcommand{\DNDRSS}{DN+DR+SS}
\newcommand{\CBF}{CBF}
\newcommand{\BF}{beamformer}
\newcommand{\form}{implementation}
\newcommand{\MIMO}{source-packed}
\newcommand{\Mimo}{Source-packed}
\newcommand{\SW}{target-wise}
\newcommand{\Sw}{Target-wise}
\newcommand{\target}{single-target}
\newcommand{\argmax}{\operatornamewithlimits{argmax}}